\documentclass[12pt,epsf]{article}
\usepackage{verbatim}

\usepackage{dsfont}
\usepackage{tikz}
\usepackage{sidecap}
\sidecaptionvpos{figure}{t}
\usepackage{subfigure}

\usepackage[english]{babel}
\usepackage{enumitem}  

\usepackage{amssymb,amsmath}
\usepackage{graphicx, xcolor, varwidth}
\usepackage{setspace}
\usepackage[permil]{overpic}
\usepackage{cite}

\DeclareSymbolFont{matha}{OML}{txmi}{m}{it}
\DeclareMathSymbol{v}{\mathord}{matha}{118}

\colorlet{darkblue}{blue!70!black}
\colorlet{darkgreen}{green!70!black}

\usepackage[colorlinks=true,urlcolor=darkblue,linktocpage=true,linkcolor=darkblue,citecolor=darkblue]{hyperref}

\numberwithin{equation}{section}

\newcommand{\be}{\begin{equation}}
\newcommand{\ee}{\end{equation}}
\newcommand{\bea}{\begin{eqnarray}}
\newcommand{\eea}{\end{eqnarray}}
\newcommand{\bear}{\begin{eqnarray}}
\newcommand{\eear}{\end{eqnarray}}  
\newcommand{\beas}{\begin{eqnarray*}}

\newcommand{\eeas}{\end{eqnarray*}}
\newcommand{\ba}{\begin{array}}
\newcommand{\ea}{\end{array}}

\newcommand{\nn}{\nonumber}

\newcommand{\pd}[2][1]{\ifnum#1=1 \frac{\partial}{\partial {#2}} \else
  \frac{\partial^#1}{\partial {#2}^{#1}}\fi}
\newcommand{\dpd}[2][1]{\ifnum#1=1 \dfrac{\partial}{\partial {#2}} \else
  \frac{\partial^#1}{\partial {#2}^{#1}}\fi}
\newcommand{\td}[2][1]{\ifnum#1=1 \frac{d}{d{#2}} \else
  \frac{d^#1}{d{#2}^{#1}}\fi}

\newcommand{\smallstart}{
\end{spacing}
\noindent\hfil\rule{1\textwidth}{.4pt}\hfil\small

   \addtolength{\leftskip}{5mm}
}
\newcommand{\smallend}{
   \addtolength{\leftskip}{-5mm}
\noindent\hfil\rule{1\textwidth}{.4pt}\hfil\normalsize
\begin{spacing}{1.3}
}

\newcommand{\blockS}[5]{
\begin{tikzpicture}[baseline=-3pt]
\coordinate (v1) at (-1,-0.5) {} {};
\coordinate (v2) at (-0.5,0) {} {};
\coordinate (v3) at (-1,0.5) {} {};
\coordinate (v4) at (0.5,0) {} {};
\coordinate (v5) at (1,0.5) {} {};
\coordinate (v6) at (1,-0.5) {} {};
\begin{scope}[very thick]
\draw  (v1) node[left] {$#2$}--(v2);
\draw  (v3) node[left] {$#1$}-- (v2);
\draw  (v4) -- (v2) node[midway, above] {$#5$};
\draw  (v5) node[right] {$#3$}-- (v4);
\draw  (v6) node[right] {$#4$}-- (v4);
\end{scope}
\end{tikzpicture}
}

\newcommand{\blockT}[5]{
\begin{tikzpicture}[baseline=-3pt]
\coordinate (v1) at (-0.5,1) {} {} {};
\coordinate (v2) at (0,0.5) {} {} {};
\coordinate (v3) at (0.5,1) {} {} {};
\coordinate (v4) at (0,-0.5) {} {} {};
\coordinate (v5) at (0.5,-1) {} {} {};
\coordinate (v6) at (-0.5,-1) {} {} {};
\begin{scope}[very thick]
\draw  (v1) node[left] {$#1$}--(v2);
\draw  (v3) node[right] {$#3$}-- (v2);
\draw  (v4) -- (v2) node[midway, right] {$#5$};
\draw  (v5) node[right] {$#4$}-- (v4);
\draw  (v6) node[left] {$#2$}-- (v4);
\end{scope}
\end{tikzpicture}
}

\newcommand{\nbox}{{\,\lower0.9pt\vbox{\hrule \hbox{\vrule height 0.2 cm \hskip 0.19 cm \vrule height 0.2 cm}\hrule}\,}}

\newcommand{\ie}{{\it i.e.,}\ }

\newcommand{\half}{\tfrac{1}{2}}

\newcommand{\bz}{\bar{z}}

\newcommand{\bF}{\mathcal{F}}
\newcommand{\bO}{\mathcal{O}}

\renewcommand{\O}{\mathcal{O}}



\newcommand{\sdot}{\!\cdot\!}

\newcommand{\etal}{\textit{et al}.}

\newcommand{\e}{\varepsilon}
\newcommand{\ebar}{\overline{e}}

\begin{document}
\begin{spacing}{1.3}
\begin{titlepage}

\begin{center}
{\Large \bf  Einstein gravity 3-point functions from \\ \vspace{0.5cm} conformal field theory}

\vspace*{12mm}

Nima Afkhami-Jeddi, Thomas Hartman, Sandipan Kundu,  and Amirhossein Tajdini \\

\vspace*{6mm}

\textit{Department of Physics, Cornell University, Ithaca, New York\\}

\vspace{6mm}

{\tt \small na382@cornell.edu, hartman@cornell.edu, kundu@cornell.edu, at734@cornell.edu}

\vspace*{15mm}
\end{center}
\begin{abstract}

 We study stress tensor correlation functions in four-dimensional conformal field theories with large $N$ and a sparse spectrum. Theories in this class are expected to have local holographic duals, so effective field theory in anti-de Sitter suggests that the stress tensor sector should exhibit universal, gravity-like behavior.  At the linearized level, the  hallmark of locality in the emergent geometry is that stress tensor three-point functions $\langle TTT\rangle$, normally specified by three constants, should approach a universal structure controlled by a single parameter as the gap to higher spin operators is increased. We demonstrate this phenomenon by a direct CFT calculation.  Stress tensor exchange, by itself, violates causality and unitarity unless the three-point functions are carefully tuned, and the unique consistent choice exactly matches the prediction of Einstein gravity. Under some assumptions about the other potential contributions, we conclude that this structure is universal, and in particular, that the anomaly coefficients satisfy $a\approx c$ as conjectured by Camanho \etal\ The argument is based on causality of a four-point function, with kinematics designed to probe bulk locality, and invokes the chaos bound of Maldacena, Shenker, and Stanford. 
 
\end{abstract}

\end{titlepage}
\end{spacing}

\vskip 1cm

\setcounter{tocdepth}{1}
\tableofcontents

\begin{spacing}{1.3}

\section{Towards Einstein Gravity from CFT}

AdS/CFT strongly suggests that spacetime is emergent in quantum gravity. It is an approximate, low-energy description of the fundamental microscopic degrees of freedom, and its most basic properties, including locality and gravity, arise from collective dynamics of those microscopic constituents.

Many features of gravity, from scattering to black hole entropy, can be reproduced by microscopic CFT calculations in supersymmetric gauge theory \cite{match}.  But holographic duality is likely to extend to a very large (perhaps infinite) class of CFTs, with and without supersymmetry.  The theories in this universality class differ in their microscopic details, but all produce emergent geometry at low energies, and all exhibit the rich phenomena associated to Einstein gravity in the bulk.  The mechanism responsible for this remarkable universality in a class of strongly interacting quantum field theories remains, to a large extent, mysterious.  Although specific examples can be matched in great detail, there has been no universal CFT derivation of the Bekenstein-Hawking entropy formula, the universal ratio of shear viscosity to entropy density  $\eta/s = 1/4\pi$\cite{Kovtun:2003wp}, or indeed any other prediction of Einstein gravity  in more than three bulk dimensions --- without higher curvature corrections ---  that is not fixed by conformal invariance or other symmetries. A universal explanation should not only match the correct answer, but also shed light on why it is independent of the microscopic details. 

On the gravity side, universality is guaranteed by effective field theory. At low energies, any theory of quantum gravity is described by the Einstein action, plus small corrections suppressed by the scale $M$ of new physics:
\be\label{sgrav}
S \sim \frac{1}{G_N}\int\sqrt{-g}\left(-2\Lambda + R + \frac{c_2}{M^2} R^2 + \frac{c_3}{M^4} R^3 + \cdots\right) \ .
\ee
Typically the dimensionless coefficients $c_i$ in effective field theory are order one, so that the Einstein term dominates at low energies, and the theory is local below the cutoff. This suppression of the higher curvature terms is responsible for universality -- it ensures, for example, that black hole entropy is Area$/4G_N$ up to small corrections. But if we choose the higher curvature terms to be ghost-free, then we can attempt to tune up the coefficients, $c_i \gg 1$. The theory is still weakly coupled due to the overall $1/G_N$. If this tuning is permissible, then universality may be violated, since the higher curvature terms modify the theory at energies much below the scale $M$ of new particles.

At the level of 3-graviton couplings, it was shown by Camanho, Edelstein, Maldacena, and Zhiboedov (CEMZ) \cite{Camanho:2014apa} that such tunings are, in fact, not allowed. They would lead to causality-violating propagation of gravitons in nontrivial backgrounds.  There is some question as to whether this is acceptable classically \cite{Papallo:2015rna,Andrade:2016yzc}, but in any case, it is incompatible with unitarity in the quantum theory \cite{Bellazzini:2015cra}. The conclusion is that universality of the 3-graviton coupling is a fundamental property of quantum gravity, that cannot be violated even by fine tuning. 

This poses a clear question for quantum field theory: Why, in some class of large-$N$ CFTs, must the stress tensor 3-point functions $\langle TTT\rangle$ be tuned to a particular form? In a general 4d CFT, there are three independent terms:
\be
\langle TTT\rangle_{cft} = n_s \langle TTT\rangle_s + n_f \langle TTT\rangle_f + n_v \langle TTT\rangle_v  , 
\ee
where the $\langle TTT\rangle_i$ are known tensor structures, fixed by conformal invariance,  and the $n_i$ are coupling constants \cite{Osborn:1993cr}. Einstein gravity predicts one particular structure, where only the overall coefficient is adjustable. So it imposes two relations on the three coupling constants $n_s$, $n_f$, and $n_v$ \cite{Henningson:1998gx,Nojiri:1999mh}.  One of these relations can be stated in terms of the Weyl anomaly coefficients as 
\be\label{ac}
a=c \ .
\ee

We will show how this universality arises in large-$N$ CFTs in $d=4$, under an assumption about double trace operator contributions described below. The argument is based on causality, like the gravity analysis of CEMZ. We consider the four-point correlation function 
\be
G = \langle\, \psi \, \psi  \, T_{\mu\nu} \, T_{\alpha\beta} \, \rangle
\ee
where $\psi$ is a scalar operator and $T$ is the stress tensor. In \cite{causality1,causality2,Hofman:2016awc,anec} it was shown that causality of this correlation function, in the lightcone limit, leads to the Hofman-Maldacena conformal collider bounds \cite{Hofman:2008ar}. For example, it constrains the anomaly coefficients to lie within the window $\frac{1}{3} \leq \frac{a}{c} \leq \frac{31}{18}$.  In order to derive stronger constraints such as \eqref{ac} in large-$N$ theories, we apply similar logic, but in different kinematics that are designed to probe bulk locality. Applying the chaos bound derived by Maldacena, Shenker, and Stanford \cite{mss}, we show that causality of this 4-point function in large-$N$ CFT requires carefully tuned 3-point functions. The tuning agrees precisely with the universal structure predicted by Einstein gravity, without higher curvature corrections. The connection to chaos is that the 4-point function in Minkowski spacetime can be viewed as a thermal correlator in Rindler space, and in this context, the bound on chaos follows from analyticity and can be interpreted as a causality constraint.

The analysis closely follows the derivation of the averaged null energy condition in \cite{anec}, so it is tempting to interpret $a=c$ as a generalized, holographic energy condition; this may have interesting connections to recent work on holographic entanglement \cite{Lashkari:2014kda,Lashkari:2016idm}, just as the averaged null energy condition ties together entanglement and causality in the lightcone limit \cite{Faulkner:2016mzt,anec}.

Our starting point is the assumption that there is a large gap $\Delta_{gap}$ in the spectrum of scaling dimensions for single-trace operators, since this is the case in any CFT with a nice holographic dual. Otherwise, the effective field theory in the bulk cannot be truncated to any finite number of fields. This condition, referred to loosely as a sparse spectrum, has played a key role in much of the progress on understanding universality in large-$N$ CFT over the last several years. Most of this work has focused on two areas: 3d gravity from 2d CFT (for example \cite{Strominger:1997eq,Keller:2011xi,Hartman:2013mia,Hartman:2014oaa,Fitzpatrick:2014vua,Perlmutter:2016pkf}), and, in higher dimensions, applying conformal bootstrap methods to investigate the emergence of bulk effective field theory (see in particular \cite{Cornalba:2006xk,Cornalba:2006xm,Cornalba:2007zb,Heemskerk:2009pn,Mack:2009gy,Mack:2009mi,Fitzpatrick:2010zm,Heemskerk:2010ty,Fitzpatrick:2011hu,Fitzpatrick:2011ia,
ElShowk:2011ag,Komargodski:2012ek,Fitzpatrick:2012yx,Fitzpatrick:2012cg,Goncalves:2014rfa,Hijano:2015zsa,Alday:2016htq}). The latter program has focused mostly on the correlators of scalar fields, though some results on spinning fields are also available \cite{Costa:2011mg,Costa:2014kfa}. We will rely heavily on the point of view taken in this approach, and aim to extend parts of it, at the linearized level, to the graviton.  One way to approach this would be to solve the crossing equation for spinning operators, as was done for scalars in \cite{Heemskerk:2009pn}. This would be necessary to find a detailed match to much of the gravity literature. However, for the purpose of understanding the universal 3-point functions we will find that it is not required, and take a different approach that does not rely on the details of the crossing equation. Another approach to crossing in the gravitational sector has been taken in \cite{Alday:2014tsa} using supersymmetry, which reduces it to a problem with external scalars. There are hints of similar universality, but in these theories $a=c$ is fixed by supersymmetry.

The argument for universality of $\langle TTT\rangle$ has two aspects.  First, we consider the contribution to the correlator from the exchange of the stress tensor and all its conformal descendants -- that is, the stress tensor conformal block.  In the Regge limit (as defined and studied in \cite{Brower:2006ea,Cornalba:2007fs,Cornalba:2008qf,Costa:2012cb}), we find that the stress tensor block produces contributions to the four-point function that violate the chaos bound, unless the couplings are tuned to their universal Einstein ratios.  Specifically, the limit we take is the Regge limit (in terms of conformal cross ratios, $z,\bz \to 0$) followed by a bulk point limit ($z \to \bz$). This is designed to probe local, high energy scattering deep in the bulk of the dual geometry.

The second step is to understand how this conformal block for stress tensor exchange is related to the behavior of the full correlator, so this requires understanding the contributions from other operators. If these other operators can be chosen to cancel the causality violation from stress tensor exchange, then $\langle TTT\rangle$ is not universal. We discuss the various possibilities, and to some extent rule out the possibility that causality violation can be fixed by any operators present below the gap:
\begin{itemize}
\item Low-spin operators with $\ell <2$. These do not affect the Regge limit, so can be ignored.
\item Spin-2, non-conserved operators, including double trace composites. These do appear in known examples \cite{Costa:2012cb}, and do affect the Regge limit, but we sketch an argument that they cannot be used to cancel the causality violation from stress tensor exchange in the smeared correlator if $\langle TTT\rangle$ differs from Einstein gravity. This argument is incomplete, so this is a potential loophole in the argument. Our conclusions about universality hold only under the assumption that the spin-2 double traces are indeed subleading in the smeared correlator.\footnote{This loophole was subsequently closed in \cite{Afkhami-Jeddi:2017rmx}.}  
\item Higher spin double trace operators. If present, these would also affect the Regge limit. However we use the chaos bound and crossing to show that they cannot appear at this order.
\end{itemize}
The final option is to add new single trace operators above the gap.  There must be an infinite number of such operators with arbitrarily high spin $\ell \to \infty$ in order to be compatible with the chaos bound.  In this case, we find that the constraints on the 3-point function must still hold to a good approximation, but not exactly; for example the bound on the anomaly coefficients becomes 
\be
\left| \frac{a-c}{c} \right| \lesssim \Delta_{gap}^{-2} \ .
\ee
This estimate agrees with the gravity result of CEMZ \cite{Camanho:2014apa}. It is an estimate rather than a strict bound, and it relies on the assumption that OPE coefficients of very heavy operators are such that they can be ignored at the onset of the Regge regime. It would be nice to prove this and to find a strict bound, as well as to understand the relationship to the Regge intercept. Perhaps combining the methods here with \cite{Costa:2012cb} would prove useful.

We begin by reviewing well known results about the structure of correlation functions in large-$N$ CFT in $d>2$ dimensions in section \ref{s:opereview}. In section \ref{s:reggesetup} we review the Regge limit, the bulk point limit, and the chaos bound, then setup the kinematics that will be used to extract constraints. In section \ref{s:texchange} we compute the conformal block for stress tensor exchange and evaluate the contribution in the Regge-bulk-point kinematics.  The discussion of universality, and to what extent the causality-violating contributions can be cancelled by other operators, is in section \ref{s:restored}.

\section{Conformal block expansion at large $N$}\label{s:opereview}

This section is a brief introduction to conformal block methods in large-$N$ CFT. It is entirely review, mostly following \cite{Heemskerk:2009pn,Heemskerk:2010ty}.

\subsection{Crossing}

In a conformal field theory, $n$-point correlation functions can be reduced to lower-point functions by successive application of the operator product expansion (OPE) on pairs of operators. By applying the OPE twice, a 4-point function can be decomposed as a sum on conformal blocks. Consider external scalars, identical in pairs:
\be\label{gfour}
G =  \langle \O(x_1)\mathcal{O}(x_2)\psi(x_3)\psi(x_4)\rangle \ .
\ee
The conformal block expansion is
\begin{align}\label{crossing}
G &=\frac{1}{x_{12}^{2\Delta_O}x_{34}^{2\Delta_\psi}}\sum_{p} c_{OOp} c_{\psi\psi p}g_{\Delta_p,\ell_p}(z,\bz)\notag\\
&=\frac{1}{(x_{14}x_{23})^{\Delta_O+\Delta_\psi}}\sum_{q}c_{O\psi p}^2 g_{\Delta_q,\ell_q}(1-z, 1-\bz),
\end{align}
where the sums are over primaries which appear in the OPE of both pairs of operators, and the conformal block $g_{\Delta,\ell}(z,\bz)$ contains contributions from the exchange of the primary with scaling dimension $\Delta$ and spin $\ell$ as well as all of its descendants. The coefficients $c_{ijk}$ are the couplings that appear in three-point functions $\langle \O_1 \O_2 \mathcal{O}_p\rangle$, and the conformal cross-ratios $z$ and $\bz$ are defined by
\begin{align}\label{crossratios}
&z \bz =\frac{x_{12}^2x_{34}^2}{x_{13}^2x_{24}^2}\quad\text{and}\quad (1-z)(1-\bz)=\frac{x_{14}^2x_{23}^2}{x_{13}^2x_{24}^2}.
\end{align}	
The second equality in \eqref{crossing}, called the crossing equation, follows from two different ways of applying the OPE. 
 The conformal blocks are commonly depicted as
\be
g_{\Delta_p, l_p}(z,\bz) = \blockS{\O}{\O}{\psi}{\psi}{\O_p} , \qquad
g_{\Delta_q,\ell_q}(1-z, 1-\bz) = \blockT{\O}{\O}{\psi}{\psi}{\O_q} \ .
\ee
The large-$N$ expansion provides a useful perturbative framework at strong coupling. Typically the parameter $N$ arises from considering non-abelian gauge theories with a large number of colors. Gauge invariant operators in these theories are obtained by tracing over the gauge group indices, so they can be classified as single trace, double trace, and so on. In the large-$N$ limit, correlation functions of single trace operators factorize into 2-point functions at leading order:
\begin{align}\label{deco}
\langle \Phi_1\Phi_2...\Phi_n\rangle= \langle \Phi_1\Phi_2\rangle...\langle \Phi_{n-1}\Phi_n\rangle+\text{permutations}
+\bO(1/N).
\end{align}
In this section we assume that all 2-point functions are normalized to unity, $\langle \O_i \O_j\rangle \sim \delta_{ij}$, including the stress tensor, since this is most convenient for large-$N$ counting. In the rest of the paper, the stress tensor has canonical normalization $\langle TT \rangle \sim N^2$. The coefficient in the canonically normalized stress tensor 2-point function is sometimes denoted $c_T$ and referred to as a central charge; in four dimensions, it is proportional to the Weyl anomaly $c$, so $c_T \sim c \sim N^2$.
  
  For our purposes, the gauge theory origin of $N$ plays no role, so we view \eqref{deco} as the definition of a large-$N$ theory, and adopt the nomenclature `single trace' to refer to this special class of operators. At leading order, 3-point functions of single trace operators vanish.  Double trace operators are constructed schematically as
\be
[\Phi_i\Phi_j]^{n,\ell}_{\mu_1...\mu_l} \sim \Phi_i (\Box)^n \partial_{\mu_1}...\partial_{\mu_l}\Phi_j + \cdots
\ee
where the dots are similar terms chosen to make a conformal primary. 
At large $N$, the scaling dimension of this composite operator is
\begin{align}
\Delta_{n,\ell}=\Delta_i+\Delta_j+2n+l+\gamma_{n,\ell} ,
\end{align} 
where the anomalous dimension $\gamma_{n,\ell}$ is $\sim 1/N^2$. 

The conformal block expansion and the crossing equation can be organized order by order in $1/N$. 
At leading order, \eqref{gfour} factorizes,
\be
G = \frac{1}{x_{12}^{2\Delta_O}x_{34}^{2\Delta_\psi}} + O(1/N^2) \ .
\ee
In the channel $\O\O \to \psi\psi $ this is simply the contribution of the identity operator.  Other operator exchanges contribute at $O(1/N^2)$. The large-$N$ counting for OPE coefficients is
\begin{align}
\langle \psi \psi [\psi \psi]^{n,\ell}\rangle \sim 1 \ , \qquad
\langle \psi \psi \Phi\rangle \sim 1/N \ , \qquad
\langle \psi \psi [\O\O]^{n,\ell}\rangle \sim 1/N^2
\end{align}
where $\Phi$ is a different single trace operator. Higher traces can be ignored at this order. Therefore the leading and subleading contributions, suppressing coefficients, are
\begin{align}\label{gleads}
G &= \blockS{\O}{\O}{\psi}{\psi}{1} + \sum_{\Phi}  \blockS{\O}{\O}{\psi}{\psi}{\Phi}\\
 &\qquad \qquad  \qquad + \sum_{n,\ell}  \blockS{\O}{\O}{\psi}{\psi}{[\psi\psi]^{n,\ell}}
  + \sum_{n,\ell}  \blockS{\O}{\O}{\psi}{\psi}{[\O\O ]^{n,\ell}} + O(1/N^4) \ , \nn 
\end{align}
where the first term is $\sim N^0$ and the others are $\sim 1/N^2$.
Since the double trace OPE coefficients already contribute a factor of $1/N^2$, the anomalous dimensions can be dropped in this channel.

In the other channel, $\O\psi \to \O\psi$, there is no identity exchange. Instead, the leading term must be accounted for by the exchange of $[\O \psi]$ double trace operators. At subleading order, there are three types of terms: single trace exchanges, corrections to the OPE coefficients of $\langle \O \psi [\O\psi]^{n,\ell}\rangle$ that already appeared at leading order, and anomalous dimensions of these double trace operators:
\begin{align}\label{gleadt}
G &= \sum_{n,\ell} (P_{n,\ell} + \delta P_{n,\ell} + \gamma_{n,\ell}\partial_\Delta)\blockT{\O}{\O}{\psi}{\psi}{[\O\psi]^{n,\ell}} +\sum_\Phi P_{\Phi} \blockT{\O}{\O}{\psi}{\psi}{\Phi} \quad   + \quad O(1/N^4)\ ,
\end{align}
where the $P$'s capture the leading and subleading OPE coefficients, and the block is evaluated at the canonical scaling dimension.  Again the first term is $\sim N^0$ and the others are $\sim 1/N^2$.

Comparing \eqref{gleads} to \eqref{gleadt} gives the crossing equation to order $1/N^2$. If we assume that $\O$ and $\psi$ are the only important single trace operators, and choose the dimension of $\psi$ to be large so that $[\psi\psi]$ can be ignored, then this equation becomes
\be\label{crossingat}
\blockS{\O}{\O}{\psi}{\psi}{\O} + \sum_{n,\ell}  \blockS{\O}{\O}{\psi}{\psi}{[\O\O]^{n,\ell}}
 = \sum_{n,\ell} (\delta P_{n,\ell} + \gamma_{n,\ell}\partial_\Delta)\blockT{\O}{\O}{\psi}{\psi}{[\O\psi]^{n,\ell}}  \ .
\ee

The story is similar if the external operators $\O$ carry spin.  In this case, there are multiple tensor structures in the 3-point function $\langle \O \O \mathcal{O}_p\rangle$, with independent coefficients, $c_{\O\O p}^{k=1,2,\dots}$. Since the full conformal block has two 3-point vertices, it is a sum of terms proportional to products of OPE coefficients, $c_{\O \O p}^j c_{\O \O p}^k$ \cite{Costa:2011mg,Costa:2011dw}. Similarly, there are several different types of double trace composites $[\O\O]^{n,\ell}$, corresponding to the different representations that can appear in the tensor product. The overall structure of the crossing equation, and classification of terms that appear at $O(1/N^2)$, is otherwise the same.

We will take $\O$ to be the stress tensor $T_{\mu\nu}$, and mostly focus on the contributions to the left-hand side of \eqref{crossingat}. The crossed channel on the right-hand side will only be used indirectly to bound the contribution of high spin double trace operators on the left.

\subsection{Sparseness and universality}

In a holographic CFT, each low-dimension single trace primary corresponds to a bulk field.  If we expect the holographic dual to have an ordinary effective field theory description, then we must assume that the number of such primaries is finite in the $N \to \infty$ limit. This is referred to as a `sparse spectrum.' In 2d CFT, it has played an essential role in classifying holographic CFTs and understanding emergent universality \cite{Hartman:2014oaa}. 

In the two-dimensional context, the definition of sparse is actually rather loose: we only require the low-lying density of states to be sub-Hagedorn, $\rho(\Delta) \lesssim e^{2\pi \Delta}$. An intuitive way to understand why even an exponentially large number of light operators have universal behavior is to note that the gravitational sector in 3d is fixed by Virasoro symmetry up to an overall coefficient, the central charge. Higher curvature corrections can therefore be absorbed into a field redefinition, so stringy corrections to the gravity action play little role.  

In higher dimensions, universality is much more subtle.  (See for example \cite{Belin:2016yll} for a recent discussion.) The gravitational action has higher curvature corrections that depend on the string tension, and cannot be absorbed into a field redefinition. Therefore, to find universal behavior in CFT, we must impose a stronger restriction on the spectrum, to ensure that the dual string states are heavy. In \cite{Heemskerk:2009pn}, it was conjectured that universality is achieved in theories where the dimension of the lightest single-trace operator of spin $\ell > 2$ has dimension
\be
\Delta_{gap} \gg 1 \ .
\ee
We adopt this as our definition of sparse. 
Our goal is to understand how universality in the (linearized) gravity sector arises in the $\Delta_{gap} \to \infty$ limit, and how non-universal corrections are suppressed at finite but large $\Delta_{gap}$.

\section{Sign constraints in the Regge limit}\label{s:reggesetup}

\begin{figure}

\begin{center}
\usetikzlibrary{decorations.markings}    
\usetikzlibrary{decorations.markings}    
\begin{tikzpicture}[baseline=-3pt]

\begin{scope}[very thick,shift={(-4,0)}]
\coordinate (v1) at (-1.5,-1.5) {};
\coordinate(v2) at (1.5,1.5) {};
\coordinate (v3) at (1.5,-1.5) {};
\coordinate(v4) at (-1.5,1.5) {};

\draw[thin,-latex]  (v1) -- (v2)node[left]{$v$};
\draw[thin,-latex]  (v3) -- (v4)node[right]{$u$};
\draw  (0,3) -- (3,0);
\draw  (0,3) -- (-3,0);
\draw  (0,-3) -- (3,0);
\draw  (0,-3) -- (-3,0);
\coordinate(v5) at (0,3) {};
\draw(v5)node[above]{Lightcone Limit};
	\draw[blue,thick,->] (-1.2,.6)--(-.9,.9);
	\draw[blue,thick,->] (1.2,-.6)--(.9,-.9);
	\draw[fill,black] (1.2,-.6) circle (0.04) node[right,black]{$\psi$};
	\draw[fill,black] (-1.2,.6) circle (0.04) node[left,black]{$\psi$};

	\draw[fill,black] (1,0) circle (0.04) node[right,black]{$T$};
	\draw[fill,black] (-1,0) circle (0.04) node[left,black]{$T$};
\end{scope}

\begin{scope}[very thick,shift={(4,0)}]
\coordinate (v1) at (-1.5,-1.5) {};
\coordinate(v2) at (1.5,1.5) {};
\coordinate (v3) at (1.5,-1.5) {};
\coordinate(v4) at (-1.5,1.5) {};

\draw[thin,-latex]  (v1) -- (v2)node[left]{$v$};
\draw[thin,-latex]  (v3) -- (v4)node[right]{$u$};
\draw  (0,3) -- (3,0);
\draw  (0,3) -- (-3,0);
\draw  (0,-3) -- (3,0);
\draw  (0,-3) -- (-3,0);
\coordinate(v5) at (0,3) {};
\draw(v5)node[above]{Regge Limit};
\def \fac {.6};
\draw[
	scale=.5,samples=50,thick,blue,domain=2.5:4.6,variable=\y,
	postaction=decorate, 
	decoration={markings, 
		mark=at position 0 with {\draw[fill,black] circle (0.04) node[left,black]{$\psi$};},
		mark=at position 1 with {\arrow{>}},
		}]
	plot ({-\fac/(\y)-\fac*\y},{-\fac/(\y)+\fac*\y});
\draw[
	scale=.5,samples=50,thick,blue,domain=2.5:4.6,variable=\y,
	postaction=decorate, 
	decoration={markings, 
		mark=at position 0 with {\draw[fill,black] circle (0.04) node[right,black]{$\psi$};},
		mark=at position 1 with {\arrow{>}},
		}]
	plot ({\fac/(\y)+\fac*\y},{\fac/(\y)-\fac*\y});
	\draw[fill,black] (1,0) circle (0.04) node[right,black]{$T$};
	\draw[fill,black] (-1,0) circle (0.04) node[left,black]{$T$};
\end{scope}

\end{tikzpicture}
\end{center}
\caption{\label{fig:regge} \small The lightcone limit is $v \to 0$ with $u$ fixed.  The Regge limit is $v \to 0$ with $uv$ fixed. In general, these limits do not commute.}
\end{figure}
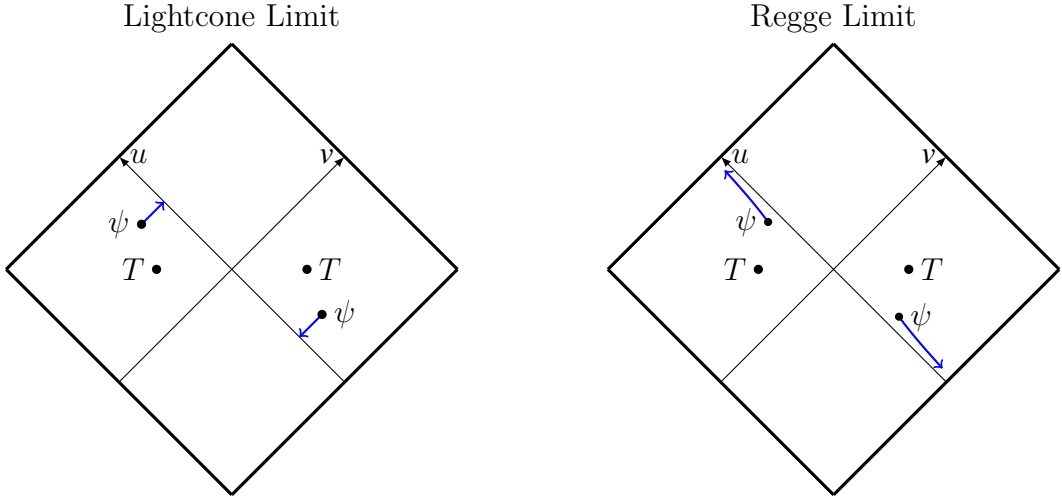

Consider the CFT correlation function
\be\label{reggeG}
G_{local} = \langle  \psi(x_3) \e_1 \sdot T(x_1) \e_2 \sdot T(x_2) \psi(x_4) \rangle
\ee
where $T_{\mu\nu}$ is the stress tensor and $\psi$ is a scalar primary. The $\e$'s are null polarization vectors,  and $\e \sdot O \equiv \e^\mu \e^\nu \cdots O_{\mu\nu\cdots}$. The Regge regime is a Lorentzian limit of the 4-point function with $z,\bz \to 0$, holding the ratio $z/\bz$ fixed \cite{Brower:2006ea,Cornalba:2007fs,Cornalba:2008qf,Costa:2012cb}. 
One way to reach it is to set\footnote{All points and polarization vectors are written in Lorentzian components $(t, y, \vec{x})$; Euclidean time is $\tau = it$.}
\begin{align}
x_1 &= (0, 1, \vec{0})  &x_2 = -x_1\\
x_3 &= (t, y, \vec{0})  &x_4 = -x_3   \nn
\end{align}
then take 
\be
v \to 0 ,\quad u \to \infty , \quad uv \ \mbox{fixed} \ ,
\ee
where $u = t-y , \  v = t + y$. This is illustrated in figure \ref{fig:regge}. The operator ordering in \eqref{reggeG} is also important, since it specifies what path to take in the complex $z$ and $\bz$ planes in order to reach the Regge regime.  The correct choice, with the ordering \eqref{reggeG}, is to send $z$ around the singularity at $z=1$ before taking $z,\bz \to 0$ \cite{Cornalba:2007fs,Cornalba:2008qf}. See \cite{causality1} for a general discussion of ordering and analytic continuation in Lorentzian QFT.

Let us set $\bz = \eta z$, so that $z \to 0$ at fixed $\eta$ is the Regge limit.  Also suppose the two polarizations are related by a Rindler reflection:
\be
\e_1 = \overline{\e}_2 , \quad \overline{V} \equiv (-V^t, -V^y, \vec{V}\, )^* \ .
\ee
Maldacena, Shenker and Stanford \cite{mss} showed that correlators of this type that behave in the Regge regime as
\be\label{chaosc}
G_{local} \propto 1 + \frac{1}{N} \delta G_{local} , 
\ee
where $1/N$ is small parameter, obey two constraints. First, they can grow no faster than the power $\delta G_{local} \sim 1/z$ as $z\to 0$, and second, Im $\delta G_{local} < 0$ (for real $z,\bz$). This is known as the chaos bound, due to Kitaev's interpretation of this growth as Lyapunov behavior \cite{Kitaev:talk}. It can also be interpreted as a causality constraint \cite{Camanho:2014apa,mss}. Although we inserted the local operator $T(x_1)$, the bound also applies when the $T$ insertions are smeared in the left and right Rindler wedges, with the smearing functions related by a Rindler reflection.

The Regge version of the chaos bound requires large $N$, so that the correlator in the Regge regime can be separated into the identity contribution plus a small correction, as in \eqref{chaosc}. Similar logic was adapted to general CFTs in \cite{causality1} by taking the lightcone limit, $\eta \to 0$ with $z$ fixed, instead of the large $N$ limit. This can be done in any CFT and leads to a derivation of the averaged null energy condition \cite{anec}. Although the sign constraints follow the same logic as the chaos bound, the physics is different, as the lightcone and Regge regimes are generally controlled by different sets of operators. The lightcone limit is governed by operators of minimal twist, while the Regge limit is dominated by high spin exchanges. Even in large $N$ theories, the limits do not generally commute, so the corresponding constraints are inequivalent.  For example, in theories with a stringy holographic dual, the lightcone growth is controlled by the stress tensor, but the Regge growth is softened by string exchange. On the other hand in a holographic CFT with an effective Einstein gravity dual, both limits are controlled by stress tensor exchange. (This is discussed for example in \cite{Costa:2012cb,Camanho:2014apa,mss}.)

\subsection{Wavepackets and the Regge-Bulk-Point double limit}
Applying the bound with position-space insertions produces constraints on 3-point functions, but they are not typically the strongest constraints that can be extracted from this logic. In \cite{anec}, a method was found to smear the $T$ insertions in a way that automatically produces optimal constraints in the lightcone limit.  Our strategy here will be to apply the same smearing in the Regge limit, in order to extract the optimal constraints from the chaos bound. As described in \cite{anec}, the idea is to insert wavepackets centered at imaginary times,
\be
x^{center} = (i B, 1, \vec{0}) \ , \qquad B>0 \ .
\ee
The corresponding smeared operator is
\be
T_{\mu\nu}(B) \equiv \int d\tau d^{d-2}\vec{x} e^{-( (\tau-B)^2 + \vec{x}^2)/D^2} T_{\mu\nu}(i\tau, 1, \vec{x}) \ ,
\ee
where $D$ is the width of the wavepacket. We could instead conformally map the wavepacket center to a real point in Minkowski spacetime and the analysis would be identical, but the smearing procedure would be more complicated. We will see that the wavepacket centers have real cross ratios, so this can be viewed as scattering of ordinary wavepackets.
The Rindler reflection of this operator is
\be
\overline{ \e \sdot T(B) } = \int d\tau d^{d-2}\vec{x} e^{-((\tau-B)^2 + \vec{x}^2)/D^2} \overline{\e}\sdot T (i \tau, -1, \vec{x}) \ .
\ee
Note that the imaginary time value is unchanged under Rindler reflection, which maps $t \to -t^*$ \cite{anec}. 

Thus we are led to study the correlator
\be\label{gdef}
G = \frac{1}{\mathcal{N}} \, \langle\,  \psi(t,y) \, \overline{ \e \sdot T(B) } \,  \e \sdot T(B) \, \psi(-t,-y) \, \rangle
\ee
with $\mathcal{N} \equiv \langle \psi(t,y)\psi(-t,-y)\rangle \langle \overline{ \e \sdot T(B) }  \e \sdot T(B) \rangle  >0$. The transverse arguments of $\psi$ are set to $\vec{x} = 0$. As shown in \cite{anec}, this correlator satisfies all of the assumptions of the chaos/causality bounds, for $|t|<|y|$ so that the $\psi$'s are inserted in the Rindler wedges. To take the Regge limit, first let
\be\label{uvrel}
u = 1/\sigma , \qquad v = -\sigma B^2 \rho \ ,
\ee
with $\sigma > 0$, $B>0$, $\rho \in (0,1)$. Then the Regge limit is $\sigma \to 0$ with $B,\rho$ fixed. Next, we send $B \to \infty$. This choice was motivated in \cite{anec} via the Hofman-Maldacena \cite{Hofman:2008ar} kinematics, and reflects the holographically dual AdS calculations, where probes are typically inserted far from the shockwave.  The resulting smeared correlator is a function of $\sigma$ and $\rho$, satisfying
\be
0 < \sigma \ll 1 \ , \qquad 0 < \rho < 1 \ .
\ee
$\sigma$ controls the distance from the Regge singularity. To give a physical interpretation to $\rho$, consider the conformal cross ratio at the centers of the two wavepackets, \ie
\be
x_1 = x^{center} = (iB,1) , \quad x_2 = \overline{x}^{center} = (iB,-1) , \quad x_3 = (\half(v+u), \half(v-u)) , \quad x_4 = -x_3 \ ,
\ee
with transverse coordinates $\vec{x}$ set to zero.
These points, in the limit just described, have cross ratios
\be
z = 4 \sigma , \quad \bz = 4 \rho \sigma \ .
\ee
From this we can understand the meaning of $\rho$. As $\rho \to 0$, we enter the lightcone regime $\bz \ll z \ll 1$. In this limit we expect to recover the Hofman-Maldacena bounds as in \cite{anec}. As $\rho \to 1^-$, the wavepacket centers approach the double limit
\be
z,\bz \to 0 , \qquad z \to \bz \ .
\ee
The first is the Regge limit, and the second is the `bulk point' limit, which gets its name from the fact that in holographic theories, it probes scattering at a point deep in the interior of AdS \cite{Heemskerk:2009pn,Susskind:1998vk,Polchinski:1999ry,Gary:2009ae,
Gary:2009mi,Giddings:1999jq,Maldacena:2015iua}. This limit is our main interest and it is here that we expect to recover the CEMZ causality constraints on $\langle TTT\rangle$.\footnote{Note, however, that this is not exactly the bulk point singularity studied in \cite{Maldacena:2015iua}, which in our notation would lie at negative $z,\bz$. An advantage of studying the $z\sim \bz$ enhancement at $z,\bz > 0$ is that the conformal block expansion is under better control, and in fact converges absolutely in the crossed channel \cite{causality1}.}

\subsection{The chaos bound}
We will apply the chaos bound to these wavepackets.  That is, we plug the kinematics \eqref{uvrel} into \eqref{gdef}, subtract the disconnected piece, and take the limits:
\be\label{dgdef}
\delta G(\e; \rho) \equiv \lim_{B\to \infty} \lim_{\sigma \to 0} \sigma  (G-1) \ .
\ee
(Appendix \ref{app:integrals} explains how to calculate the dominant contributions to the integrals in a way that avoids integrating the Gaussian smearing factors.) The chaos bound \cite{mss} is a general bound on certain correlations in a thermal quantum system. It actually consists of two distinct inequalities: First, the correlations cannot change faster than a certain rate, and the second, the sign of the leading correction is fixed.  Vacuum correlators in Minkowski spacetime, with insertions restricted to the Rindler wedges, have a thermal interpretation, so the chaos bound applies.   In this context, the bound on the growth implies that the $\sigma\to 0$ limit in \eqref{dgdef} is finite, and the sign constraint imposes
\be\label{fcbound}
\mbox{Im}\ \delta G \leq 0 \ .
\ee
At this stage, the parameter $\rho \in (0,1)$ is arbitrary, but as mentioned above, we will ultimately be interested in the bulk-point limit $\rho \to 1^-$. The chaos bound \eqref{fcbound} is interpreted as a causality constraint because if it is violated, then the 4-point correlator is non-analytic in a regime where it should be analytic, and this leads to non-vanishing commutators for spacelike separated operators (see \cite{causality1} for further discussion).

\section{Contribution of stress tensor exchange}\label{s:texchange}
A correlation function of local primary operators can be expanded in a sum of conformal blocks. At order $1/N^2$, 
\be
\langle \psi(x_3) T(x_1) T(x_2) \psi(x_4) \rangle =
\blockS{T}{T}{\psi}{\psi}{1}
+
\blockS{T}{T}{\psi}{\psi}{T}
+ \mbox{\small \ double-trace\ }
 + \mbox{\small \ other} \ .
\ee
As explained in section \ref{s:opereview}, the double trace contributions are composites $[TT]$ and $[\psi\psi]$. The `other' terms consist of low-spin single trace operators, and heavy single trace operators above the gap.  In this section, we will calculate the contribution of the stress tensor operator $T_{\mu\nu}$ running in the intermediate channel, evaluate it in the Regge limit, and show that generically it violates the chaos/causality bound \eqref{fcbound}. This contribution is causal only if the coupling constants in $\langle TTT\rangle $ take special values, namely those predicted by Einstein gravity in AdS$_5$. There are, however, other contributions in the Regge limit; the question of whether the stress tensor exchange actually dominates, or whether the causality-violating behavior can be cancelled by other operators, is postponed to the next section.

To understand the contribution of stress tensor exchange, we first need the 3-point functions $\langle T\psi\psi\rangle$ and $\langle TTT\rangle$. The first is completely fixed by conformal invariance, with overall coefficient $c_{T\psi\psi} \sim -\Delta_\psi/\sqrt{c}$. Here $c \sim N^2$ is the positive constant that appears in $\langle TT\rangle \sim c$.   The 3-point function of three stress tensors has three possible tensor structures in $d=4$ \cite{Osborn:1993cr}. We choose a basis of structures corresponding to free scalars, free fermions, and free vectors, so that in complete generality,
\be
\langle TTT\rangle = n_s \langle TTT\rangle_{scalar} + n_f \langle TTT\rangle_{fermion} + n_v \langle TTT\rangle_{vector} \ ,
\ee
where $n_s$, $n_f$, and $n_v$ are three numerical coupling constants (generally non-integer).
Our conventions for the structures can be found in the appendices of \cite{causality2}. We adopt all of the normalizations and notation for conformal correlators and conformal blocks described there.

The conformal block for $TT\to T \to \psi\psi$ has three independent contributions, corresponding to the three structures in $\langle TTT\rangle$. These are calculated following \cite{Costa:2011mg,Costa:2011dw}. Details and the final result in the Regge limit are in appendix \ref{app:reggeblock}. The result, for the contribution of stress tensor exchange to the correlator
\be
G_{local} = \langle \psi(x_3) \e \sdot T(x_1) \overline{\e} \sdot T(x_2) \psi(x_4)\rangle \ ,
\ee
takes the form
\be\label{glocalt}
G_{local}|_{T} = \frac{1}{(x_{34})^{2\Delta_\psi}} \frac{1}{(x_{12})^{12}} c_{\psi\psi T} \mathcal{F}_T(\e_i, x_i, z, \bz) \ ,
\ee
where $\mathcal{F}_T$ is the conformal block given in the appendix. The coupling constants $n_{s,f,v}$ have been included in the definition of the block. Unlike the scalar conformal block, $\mathcal{F}_T$ has explicit dependence on the $x_i$ via the tensor structures. 

At this stage, it is convenient to choose the null polarization
\begin{align}\label{specialep}
\e &= (1,1,i \lambda, \lambda)\\
\overline{\e} &= (-1, -1, -i \lambda, \lambda)  \ , \nn
\end{align}
where $\lambda$ is an arbitrary real number.
This makes it particularly easy to read off the positivity conditions, as we will see below. Next, we do the integrals described in section \ref{s:reggesetup}. Details are outlined in appendix \ref{app:integrals}. The final result, dropping an overall positive constant, is
\small
\begin{align}\label{dgfinal}
\delta &G = -\frac{i \rho}{\sigma(1-\rho)^5}\left[
5 n_s(1 + \rho^2) - 60 \rho n_f + (n_s + 6 n_f + 12 n_v)\rho^2(5-4\rho + \rho^2)\right]\\
&- \frac{i \rho \lambda^2}{\sigma(1-\rho)^5} \left[
- 40 n_s \rho  + 10 n_f(9+6\rho + 9\rho^2) + 2(n_s + 6n_f + 12n_v)\rho(-10 + 15 \rho - 12 \rho^2 + 3 \rho^3) 
\right]\nn\\
&- \frac{i\rho\lambda^4}{\sigma(1-\rho)^5}\big[
-5(n_s+6n_f)(3-8\rho+3\rho^2)-120\rho n_f \nn\\
& \hspace{5cm} + (n_s + 6n_f + 12n_v)(15-40\rho+45\rho^2-24\rho^3+6\rho^4)\big] \ .\nn
\end{align}
Now let us ask under what conditions this contribution to $\delta G$ obeys the chaos bound. Although \eqref{dgfinal} is for the special polarization \eqref{specialep}, it actually encodes the full set of positivity conditions on $\delta G$: we simply require each power $\lambda^0, \lambda^2$, and $\lambda^4$ to individually satisfy Im $ \delta G \leq 0$. This ensures that Im $\delta G$ is non-positive for arbitrary (not necessarily null) symmetric tensor polarizations, as shown in appendix \ref{app:nicepol}.

Expanding in the lightcone limit $\rho \to 0$, the analysis reduces to the case considered in \cite{anec}. The correlator is 
\be
\delta G \sim -i\frac{\rho}{\sigma}(n_s + 18 n_f \lambda^2 + 36 n_v \lambda^4) \ .
\ee
The positivity conditions agree with the Hofman-Maldacena conformal collider bounds \cite{Hofman:2008ar}: $n_s \geq 0$, $n_f \geq 0$, $n_v \geq 0$.  These constraints were previously derived by deep inelastic scattering in \cite{Komargodski:2016gci}, by lightcone bootstrap methods related to the present analysis in \cite{causality2,Hofman:2016awc}, and from monotonicity of relative entropy in \cite{Faulkner:2016mzt}. In this limit, the method we have just described is identical to that in \cite{anec}.

In the bulk point limit, it is more convenient to write the answer in a different basis of 3-point functions. Define
\be
t_0 = n_s + 6 n_f + 12 n_v , \quad t_2 = n_f - 2n_v , \quad t_4 = n_s - 4 n_f + 2 n_v \ .
\ee
In AdS$_5$, the corresponding structures in $\langle TTT\rangle$ are those produced by the Einstein term, the Gauss-Bonnet term, and the $R^3$ correction, respectively (also note $t_0 \propto c$). Taking $\rho \to 1^-$, \eqref{dgfinal} becomes
\begin{align}
\delta G \sim
-&\frac{i t_4}{\sigma(1-\rho)^5}(1-4\lambda^2 + \lambda^4)(1 + O(1-\rho))\\
&-\frac{5 i t_2}{6\sigma(1-\rho)^3}  (1-3\lambda^2+6 \lambda^4)(1 + O(1-\rho))\nn\\
&-\frac{i t_0}{12\sigma(1-\rho)}  (1 + 6 \lambda^2 + 6 \lambda^4)(1 + O(1-\rho))\nn
\end{align}
In the leading term, the coefficients of $\lambda^0$, $\lambda^2$, and $\lambda^4$ cannot all be positive, so this obeys the chaos bound only if
\be\label{t4bound}
t_4 = n_s - 4 n_f + 2n_v = 0 \ .
\ee
Then the $(1-\rho)^{-3}$ term dominates, and this imposes
\be\label{t2bound}
t_2 = n_f - 2 n_v = 0 \ .
\ee
Assuming that these causality-violating terms from stress tensor exchange are not cancelled by other operators (see below), it follows that the stress tensor 3-point functions $n_s$, $n_f$, and $n_v$ must obey these two constraints, $t_4 = t_2 = 0$. Only one free parameter remains, and it is fixed by the Ward identity in terms of the coefficient of the stress tensor 2-point function.  So in fact there are no free parameters in the stress tensor 3-point function.

The quantities $t_2$ and $t_4$ are proportional to the couplings of the same names in the conformal collider literature \cite{Hofman:2008ar}. Setting them to zero means that the bulk contains only the Einstein term.  Therefore our results are in perfect agreement with the gravity analysis of \cite{Camanho:2014apa}. The $\delta G$ in \eqref{dgfinal} has the same format as the time delay computed there, though we are working in the different gauge; the translation between gauges is in appendix \ref{app:nicepol}.

In four dimensions, the anomaly coefficients $a$ and $c$ are related to the couplings in $\langle TTT\rangle$. Their ratio is
\be
\frac{a}{c} = \frac{n_s + 11n_f + 62 n_v}{3(n_s + 6 n_f + 12 n_v)} \ .
\ee
Therefore setting $t_2= t_4 = 0$ also imposes 
\be
a = c \ .
\ee

\section{Discussion}\label{s:restored}

We have demonstrated that the exchange of the stress tensor conformal block leads to terms that, taken alone, violate the chaos bound.  This is a mathematical fact about conformal blocks, which in itself does not refer to any particular theory. It is a property of the conformal algebra. It is physically relevant in theories where $(i)$ this term can actually be trusted in the conformal block expansion, and $(ii)$ it is not cancelled by the contributions of other operators.  

In a typical small-$N$ CFT, the conformal block expansion breaks down before the Regge limit and we cannot trust this contribution. So in this case, there are no constraints on $t_2$, $t_4$ beyond the usual Hofman-Maldacena conditions reproduced above in the lightcone limit.  

On the other hand, in a large-$N$ theory with a sparse spectrum, the conformal block expansion is reliable even in the Regge limit, order by order in $1/N$. The basic structure was reviewed in section \ref{s:opereview}. The terms we have calculated are suppressed by $1/N^2$, and although they are enhanced by powers of $1/\sigma$ and $1/(1-\rho)$, the order of limits is such that these contributions are small everywhere we have used them. Therefore, in these theories, the contribution of the stress tensor in \eqref{dgfinal} is a reliable contribution to the correlator in a controlled expansion. How, then is causality preserved?

One simple possibility is that $t_{2,4}$ are exactly zero, as in maximally supersymmetric theories, or vanish at leading order in $1/N$.  In this case there is no need to add any new operators at this order. This corresponds to a holographic dual where the higher curvature corrections are suppressed by loop factors.  We will focus instead on the case where $t_{2,4}$ are not $1/N$ suppressed, so that the causality problem must be resolved at the same order. This corresponds to classical higher curvature terms in the bulk.

The question, then, is whether the problematic terms from stress tensor exchange can be cancelled by other operators. Low spin exchanges have no $1/\sigma$ enhancement in the lightcone limit, so these won't help. One possibility is that new single trace operators are added to the theory, and these start to dominate the correlator before the stress tensor contribution causes trouble.  We consider this scenario first, then return to (and to some extent rule out) the other possibilities below.  The discussion mirrors the analogous gravity results in \cite{Camanho:2014apa}.

\subsection{Higher spin operators}

Consider the exchange of operators with spin $\ell>2$, in addition to the stress tensor.  If there is a finite number of such operators, with maximal spin $\ell_{max}$, then the dominant Regge behavior is $\sim \sigma^{1-\ell_{max}}$.  This violates the chaos bound, so this scenario is not allowed. This argument applies to both single-trace and double-trace operators. 

So to avoid making the causality problem even worse, we must add an infinite number of higher spin exchanges, either single-trace or double-trace, with arbitrarily high spins.  We postpone the discussion of double trace operators for now, and consider the scenario where causality is saved by the exchange of an infinite tower of new, higher spin operators, not built from composites of $\psi$ and $T$.  This can certainly do the trick, since this is the expectation from string theory in the bulk. Let us estimate the size of the corrections to the formulas $t_2=t_4=0$ if we include higher spin operators with large scaling dimensions, but not large enough to compete with the $1/N$ expansion:
\be
N \gg \Delta_{gap} \gg 1 \ .
\ee
We will also assume a large gap in the twist spectrum, \ie that these higher spin operators have $\Delta \gg \ell$.
The contribution of a high dimension, high spin operator $O$ in the Regge-bulk-point limit scales as 
\be
\delta G_{O} \sim \frac{i}{\sigma^{\ell-1}} \frac{\rho^{(\Delta_{gap}-\ell)/2}}{(1-\rho)^5} \ .
\ee
The power of $\rho$ is easily checked for external scalars, and since the spinning blocks are built by acting on the scalar blocks with a fixed number of derivatives \cite{Costa:2011dw}, it appears also for external stress tensors. There are also large numerical factors in this expression, which we must assume are cancelled by small OPE coefficients. This is essentially the assumption that, even after adding these higher spin operators, the initial onset of Regge behavior is controlled by the stress tensor, rather than the higher spin operators. Then for $\rho = 1 - \epsilon$ with $\epsilon \ll 1$, 
\be
\delta G_O \sim \frac{i}{\sigma^{\ell-1} } \frac{e^{-\epsilon (\Delta_{gap}-\ell)/2} }{\epsilon^5} \ .
\ee
Therefore such operators are exponentially suppressed in the conformal block expansion, but `turn on' as we approach the bulk point limit $\epsilon \to 0$, and begin to dominate the correlator for $\epsilon \lesssim 1/\Delta_{gap}$. It follows that we cannot actually send $\epsilon \to 0$ in the argument that led to $t_4 = t_2 = 0$; instead the strongest reliable constraints come from setting $\epsilon \sim 1/\Delta_{gap}$. The form of the stress tensor contribution found above was, schematically,
\be
\delta G \sim -\frac{i}{\sigma \epsilon} \left(\pm  \frac{t_4}{\epsilon^4} \pm \frac{t_2}{\epsilon^2} + c\right)
\ee
with $c \sim n_s + 6 n_f + 12 n_v$, and the different sign choices came from different polarizations.  Therefore, setting $\epsilon \sim 1/\Delta_{gap}$ and requiring Im $\delta G \leq 0$ implies
\be\label{tconc}
\left|\frac{t_4}{c} \right| \lesssim \frac{1}{\Delta_{gap}^{4}} , \qquad \left|\frac{t_2}{c}\right| \lesssim \frac{1}{\Delta_{gap}^{2}} \ .
\ee
These scalings match the predictions of \cite{Camanho:2014apa} based on the gravity analysis, where graviton exchange is corrected by massive higher spin fields, which contribute only at very small impact parameter. The upper bound on $a-c$ is generically set by $t_2 \gg t_4$, so
\be
\left|\frac{a-c}{c}\right| \lesssim  \frac{1}{\Delta_{gap}^{2}} \ .
\ee
There are hints in the literature that even a small value of $a-c$ may have a fixed sign under certain circumstances, both from string theory examples \cite{Zwiebach:1985uq,Kats:2007mq,Blau:1999vz
,Gross:1986iv,Gross:1986mw,Tseytlin:1995bi} and from general arguments \cite{DiPietro:2014bca,Cheung:2016wjt}. Such a constraint is not evident from our analysis but it would be interesting to explore this further. Perhaps the effective field theory argument in \cite{Cheung:2016wjt}, which involves calculating the contributions from integrating out massive states, has a CFT analogue.

\subsection{Non-conserved spin-2 operators}

Now we turn to the question of whether the causality-violating contributions can be canceled by other operators already present in the light spectrum, without adding an infinite tower of higher spin states above the gap. First we consider additional, non-conserved spin-2 operators. These could be new single traces, or the double-trace operators already present in the spectrum: $[\psi\psi]^{n,2}$ and $[TT]^{n,2}$. 

Can these spin-2 operators cancel the causality-violating term? In position space, the answer is yes: double trace spin-2 operators modify the leading Regge behavior  \cite{Costa:2012cb}. However, we suspect that this is impossible after smearing the 4-point function. As mentioned in the introduction, we have not found a proof of this statement, but we will explain why it is plausible. The conclusion that the stress tensor 3-point function is universally fixed to the Einstein form holds only under the additional assumption that spin-2 double-trace operators are indeed projected out by the smearing procedure.

The rough intuition behind this assumption is that our 4-point correlator is constructed from high-momentum wavepackets. In a holographic theory, these wavepackets should effectively travel on geodesics, and the contribution to a Witten diagram from geodesic propagation has no double traces.  Clearly this argument is sensitive to the smearing procedure, and would not apply to the position-space correlator in the Regge limit, so it does not contradict \cite{Costa:2012cb}.

Finally, let us sketch one possible way to check this assumption. Non-conserved spin-2 operators couple to two stress tensors differently than conserved spin-2 operators, \ie $\langle TTX\rangle$ has a different set of tensor structures from $\langle TTT\rangle$. So we cannot expect to balance conserved operator exchange against any individual non-conserved operator exchange. In appendix \ref{app:spin8} we demonstrate this for particular examples of a single spin-2 operator.  With multiple spin-2 exchanges of various dimensions, there are more tunable parameters available, and with special choices it might be possible to restore causality in the correlator $\langle \psi\psi TT\rangle$ without setting $t_2=t_4 = 0$.  However, there may still be incurable causality violations in the correlator of four stress tensors, $\langle TTTT\rangle$. (On the bulk side, the argument that massive spin-2 particles cannot save causality also required all four external particles to be gravitons \cite{Camanho:2014apa}.) The conformal block for $TT \to T \to TT$ can be obtained by acting with derivatives on $\phi\phi \to T \to T T$ \cite{Costa:2011mg,Costa:2011dw}, so this will also have causality-violating contributions proportional to $t_2$ and $t_4$. Suppose we pick some particular polarization for two of the $T$'s  --- the two responsible for the shockwave, taking the place of $\psi$ --- and then tune the couplings of $\langle TTX\rangle$ so that causality is preserved in this particular 4-point function. We can then modify the kinematics slightly, by smearing the shockwave $T$'s or taking derivatives. This will affect the $T$ exchange and $X$ exchange differently, because $X$ is non-conserved, and therefore upset the delicate balance among coupling constants that was necessary for causality.  (An identical argument was used in the bulk \cite{Camanho:2014apa}.) To complete this argument, it would be necessary to show that smearing the shockwave does not affect the leading terms, $t_2$ and $t_4$ that appear in the constraints.

\subsection{Higher spin double trace operators}

Lastly, we return to the possibility that the infinite sum over double-trace operators $[TT]^{n,\ell}$ and $[\psi\psi]^{n,\ell}$ with $\ell >2$ cancels the problematic term from stress tensor exchange.  (In a holographic theory, the dual question is whether the results of \cite{Camanho:2014apa} can be modified by an infinite sum of contact terms, without any new propagating states.) That is, we turn on $t_{2,4} \neq 0$, and try to restore causality by adding higher spin exchanges without the introduction of any new single trace primaries.  We will argue that this is incompatible with crossing symmetry. The argument is similar to \cite{Heemskerk:2009pn} (section 7). Roughly, the idea is that the high spin double trace sum is `too analytic' to affect the Regge behavior -- it can only be modified by adding new poles or cuts to the correlator.

To satisfy the chaos bound, the sum cannot be truncated at any finite spin, so we must turn on the OPE coefficients $c_{\psi\psi[TT]}(n,\ell)$ and/or $c_{TT[\psi\psi]}(n,\ell)$ for $\ell \to \infty$. First we would like to understand under what circumstances this infinite sum can alter the Regge behavior of the correlator
\be
G(z,\bz) = \langle T_{\mu\nu}(0) T_{\rho\sigma}(z,\bz) \psi(1) \psi(\infty) \rangle \ .
\ee
(A nice solveable example where something like this occurs is 2d CFT with higher spin symmetry \cite{Perlmutter:2016pkf}.)
The sum of $[TT]$ exchanges takes the form
\be\label{ttts}
\sum_{m,n \geq 0} z^m \bz^n \left( a_{m,n}(z,\bz)   + b_{m,n}(z,\bz) \log(1-z) \right) \ ,
\ee
where the $a_{m,n}$ and $b_{m,n}$ are analytic at $z=1$ and include the tensor structures. The exchange of $[\psi\psi]$ is similar. In the Euclidean regime, each term in this sum is regular as $z,\bz \to 0$.  The Regge singularity comes from going to the second sheet, $\log(1-z) \to \log(1-z) - 2\pi i$, so it comes entirely from the log term,
\be\label{logterms}
-2\pi i \sum_{m,n \geq 0} z^m \bz^n  b_{m,n}(z,\bz) \ .
\ee
In the Regge limit $z \sim \sigma$, $\bz \sim \sigma$, $\sigma \to 0$, the contribution of a given primary is  $\sim \sigma^{1-\ell}$, so this is a sum of increasingly singular terms.  For this sum to behave as $\sim 1/\sigma$, as it must to cancel the contribution we found from the stress tensor, the sum in \eqref{logterms} must diverge at small $\sigma$.  Otherwise, it is a convergent Laurent series that can be analyzed term by term. This implies that there is a non-analyticity (a singularity or branch point) elsewhere in the $z$-plane. The convergent expansion in the other channel $T \psi \to T\psi$ requires this non-analyticity to be at $z=1$. The first term in \eqref{ttts} has the same leading behavior as the log term, so it must also be non-analytic at $z=1$. 

So the conclusion is that in order to affect the Regge behavior without violating the chaos bound, the first term in \eqref{ttts} must be non-analytic at $z=1$. But, in a theory where the only exchanges are the stress tensor and double trace operators, this would violate crossing. In the dual channel, $T\psi \to [T\psi]_{n,\ell} \to T \psi$, the sum is over non-negative, integer powers of $(1-z)$ and $(1-\bz)$. The only non-analyticity is the $\log(1-z)$ coming from anomalous dimensions.  Comparing the non-log terms in the $s$ and $t$ channels, we see that the first term in \eqref{ttts} must be analytic at $z=1$.  The comparison can be done by approaching this point from $|z| > 1$, where both sides converge.

Note that the argument in this subsection does not apply to the exchange of an infinite tower of spin-2 operators, which can satisfy the chaos bound without introducing any new non-analyticity at $z=1$. It also does not apply to an infinite tower of higher-spin single traces, since these are allowed to be non-analytic at $z=1$. Therefore it does not conflict with the results of \cite{Costa:2012cb}, showing that both spin-2 double traces and towers of higher spin single traces do indeed modify the Regge behavior.

\subsection{Summary}
To recap, we have shown that the stress tensor conformal block leads to causality-violating contributions to $\langle \psi\psi TT\rangle$, proportional to the coupling constants $t_4$ and $t_2$ that appear in $\langle TTT\rangle$.  Only the tensor structure corresponding to Einstein gravity in AdS$_5$ obeys the chaos bound.  Therefore, if $t_4$ or $t_2$ is nonzero, this contribution must be cancelled by other operators -- non-conserved spin-2, an infinite tower of high-spin double trace operators, or an infinite tower of new single trace exchanges. We gave a partial argument that non-conserved spin-2 operators cannot cure the causality violation because of the different way in which non-conserved operators couple to the stress tensor. Crossing symmetry was used to rule out a cure using high spin double trace operators.  This leaves the scenario with an infinite tower of new single trace operators above $\Delta_{gap}$, as in holographic theories with a string theory dual. Assuming that the OPE coefficients of high spin operators are small enough that we can trust our analysis, we showed that the causality constraints are relaxed, but only by terms suppressed parametrically by powers of $\Delta_{gap}$. All of these conclusions align nicely with the expectations from effective field theory in AdS$_5$, with higher curvature terms suppressed by the masses of higher spin particles.

\vspace{1cm}

\bigskip

\textbf{Acknowledgments}
We thank Diego Hofman, Sachin Jain, Zohar Komargodski, Juan Maldacena, Joao Penedones, Eric Perlmutter, David Poland, Andy Strominger, and Sasha Zhiboedov for useful discussions.  The work of NAJ, TH, and AT is supported by DOE grant DE-SC0014123, and the work of SK is supported by NSF grant PHY-1316222. The work of NAJ is also supported by the Natural Sciences and Engineering Research Council of Canada. We also thank the Galileo Galilei Institute for Theoretical Physics for providing additional travel support and where some of this work was done.

\appendix

\section{Conformal block in the Regge limit}\label{app:reggeblock}
In this appendix we give the conformal block for stress tensor exchange in the correlator
\be
\langle T(x_1) T(x_2) \psi(x_3) \psi(x_4) \rangle \ .
\ee
It is calculated on a computer using the method of \cite{Costa:2011dw}.
Many of the intermediate steps are described in \cite{causality2}; only the final limit is different (now Regge instead of lightcone). The final result is an expression consisting of tensor structures, cross ratios $z$ and $\bz$, and various combinations of the Dolan-Osborn scalar conformal blocks \cite{Dolan:2000ut} and their derivatives.  In order to approach the Regge point, these are evaluated on the 2nd sheet, \ie after sending 
\be
\log (1-z) \to \log (1-z) - 2 \pi i
\ee
in the Dolan-Osborn blocks. The conformal tensor structures appearing the correlator are defined by
\begin{align}
H_{ij}&=-2x_{ij}.\epsilon_j x_{ij}.\epsilon_i+x_{ij}^2\epsilon_i.\epsilon_j\\
V_{ijk}&=\frac{x_{ij}^2 x_{ik}.\epsilon_i-x_{ik}^2 x_{ij}.\epsilon_i}{x_{jk}^2} \ . \nn
\end{align}
With our choice of kinematics, these structures are regular in the Regge limit, and obey several relations in their leading terms so that they can be written as
\begin{align}\label{tensorexp}
V_{123}&=a+\alpha b+\bO(\alpha)^2\\
V_{124}&=a-\alpha b+\bO(\alpha)^2\nn\\
V_{213}&=f+\alpha g+\bO(\alpha)^2\nn\\
V_{214}&=f-\alpha g+\bO(\alpha)^2\nn\\
H_{12}&=h.\nn
\end{align}
Here $\alpha$ is a formal power-counting parameter, keeping track of the Regge limit. To take the limit we send $\alpha \to 0$, scaling $z \sim \alpha$ and $\bz \sim \alpha$. The result is the conformal block for stress tensor exchange in the Regge limit:
\footnotesize
\begin{align}
\mathcal{F}_T 
&= - \frac{2560 i \pi  \bz}{3 c_T z (z-\bz)^9}\bigg[(z+\bz)^2 \left(216 n_f a^2 f^2 z^2 \left(2 z^2+z \bz+2 \bz^2\right) (z-\bz)^2+864 n_v a^2 f^2 z^2 (z-\bz)^4\right) \notag \\
&\left.+3456 n_v z^2 \left(b g (h-2 a f)+a^2 g^2+b^2 f^2\right)+288 n_v \bz h \left(10 z^2-5 z \bz+\bz^2\right) (b f-a g)\right.\notag\\ 
&\left.+540 n_f a f z^2 \left(z^3+2 z^2 \bz+2 z \bz^2+\bz^3\right) (a g-b f)+216 n_f b^2 f^2 z^2 \left(3 z^2+4 z \bz+3 \bz^2\right)\right) \notag \\
&\left.+1920 n_s b g z^2 \left(2 z^2+3 z \bz+2 \bz^2\right) (b f-a g)+3456 n_v a f z^2 (z-\bz)^4 (a g-b f)\right) \notag \\
&\left.+48 b^2 f^2 \left(3 z^4+8 z^3 \bz+13 z^2 \bz^2+8 z \bz^3+3 \bz^4\right)+480 b^2 g^2 \left(3 z^2+8 z \bz+3 \bz^2\right)\right) \notag \\
&+ 4 (z-\bz)^2 \left(-6 n_s h \left(6 z^5-26 z^4 \bz+9 z^3 \bz^2-15 z^2 \bz^3+7 z \bz^4-\bz^5\right) (b f-a g)\right.\notag\\ 
&+ 4 n_s z^2 \left(a^2 f^2 \left(19 z^6+48 z^5 \bz+87 z^4 \bz^2+112 z^3 \bz^3+87 z^2 \bz^4+48 z \bz^5+19 \bz^6\right)\right.\notag\\
&+ (z+\bz) \left(8640 n_f b g z^2 (z-\bz)^2 (b f-a g)\right.\notag \\
&\left.+48 a^2 g^2 \left(3 z^4+8 z^3 \bz+13 z^2 \bz^2+8 z \bz^3+3 \bz^4\right)\right.\notag\\ 
&\left.-96 a b f g \left(7 z^4+17 z^3 \bz+22 z^2 \bz^2+17 z \bz^3+7 \bz^4\right)\right.\notag\\ 
&\left.+12 a f \left(9 z^5+23 z^4 \bz+38 z^3 \bz^2+38 z^2 \bz^3+23 z \bz^4+9 \bz^5\right) (a g-b f)\right.\notag\\ 
&+ 9 h^2 (z-\bz)^6 \left(n_f \left(3 z^2-16 z \bz+8 \bz^2\right)+8 n_v \left(3 z^2-4 z \bz+2 \bz^2\right)\right) \notag \\
&\left.+n_s a f h \left(14 z^6-56 z^5 \bz+9 z^4 \bz^2-14 z^3 \bz^3-28 z^2 \bz^4+18 z \bz^5-3 \bz^6\right)\right.\notag\\ 
&\left.-48 n_s b g z^2 h \left(2 z^2-9 z \bz+2 \bz^2\right)+216 n_f a^2 g^2 z^2 \left(3 z^2+4 z \bz+3 \bz^2\right)\right.\notag\\ 
&\left.-216 n_f a b f g z^2 \left(11 z^2+18 z \bz+11 \bz^2\right)\right.\notag\\ 
&+ (z-\bz)^4 \left(-36 n_f h \left(21 z^3-19 z^2 \bz+20 z \bz^2-4 \bz^3\right) (b f-a g)\right.\notag\\ 
&\left.+18 n_f a f h \left(15 z^4-16 z^3 \bz-5 z^2 \bz^2+16 z \bz^3-4 \bz^4\right)\right.\notag\\ 
&\left.-2160 n_f b g z^2 h+n_s h^2 \left(z^4-20 z^3 \bz+61 z^2 \bz^2-48 z \bz^3+12 \bz^4\right)\right.\notag\\ 
&\left.-144 n_v a f \bz h \left(4 z^3+5 z^2 \bz-4 z \bz^2+\bz^3\right)\right)\bigg] \ . 
\end{align}
\normalsize
$c_T$ is the normalization of the stress tensor 2-point function, following the conventions of \cite{Costa:2011mg,Costa:2011dw} (as summarized in the appendices of \cite{causality2}).

\section{Integrals}\label{app:integrals}
In this appendix we describe how to do the integrals in \eqref{dgdef}. First, in general, we can rewrite this expression in a way that avoids gaussian smearing functions by exchanging limits and integrals.  In the large-$B$ limit, one of the wavepacket integrals gives just an overall volume factor, so we can fix one of the $T$ insertions at its central value and only smear the other $T$.  Then, by taking the $B \to \infty$ limit first, we can drop the gaussians entirely, as in \cite{Hofman:2008ar}. So instead of calculating $\delta G$ as written in \eqref{dgdef}, we can   calculate
\begin{align}\label{dadef}
\delta \mathcal{A}(\e; \rho) \equiv  &\int d\tau d^2\vec{x} \lim_{B\to \infty} \lim_{\sigma \to 0} \sigma \\
&\times \frac{ \langle\,  \psi(u,v) \, \overline{\e} \sdot T(iB,-1,\vec{0}) \, \e\sdot T(i(B+\tau), 1, \vec{x})  \, \psi(-u,-v) \, \rangle_{\rm disconnected}}{ \langle \psi(u,v)\psi(-u,-v)\rangle } \nn
\end{align}
with $u,v$ set by \eqref{uvrel}. $\delta \mathcal{A}$ agrees with $\delta G$ up to a positive constant. The `disconnected' subscript indicates that we should drop the contribution of the identity conformal block, $TT \to 1 \to \psi\psi$.

In the limit $\sigma \ll 1/B \ll 1$, the conformal cross ratios are
\begin{align}
z &= \sigma\left[2(1+\rho) - i \tau(1-\rho) + \sqrt{-4\rho(4+r^2+\tau^2)-(2i+\tau+\rho(2i-\tau)^2} \right]\ , 
\\
\bz &= \sigma\left[2(1+\rho) - i \tau(1-\rho) - \sqrt{-4\rho(4+r^2+\tau^2)-(2i+\tau+\rho(2i-\tau)^2}\right] \ . \nn
\end{align}
Here we used the definition \eqref{crossratios}, with points in \eqref{dadef} labeled in the order 3,2,1,4. The tensor structures, as expressed in \eqref{tensorexp}, are 
\begin{align}
a&=-2  +i \tau +\lambda  (y_2+i y_1)\notag\\
b&=\rho  (2  -i \tau )^2-y_1^2+\lambda  (y_1-i y_2) (-2 i   (\rho +1)-\rho  \tau +\tau )-y_2^2\notag\\
f&=-2  +i \tau +\lambda  (-y_2+i y_1)\notag\\
g&=-\rho  (2  -i \tau )^2+y_1^2+\lambda  (y_1+i y_2) ((\rho -1) \tau +2 i   (\rho +1))+y_2^2\notag\\
h&=2 (2  -i \tau ) \left(2   \left(\lambda ^2+1\right)-i \left(-\lambda ^2 \tau +\tau +2 \lambda  y_1\right)\right) \ ,
\end{align}
with the transverse directions denoted $\vec{x} = (y_1,y_2)$.
The integrand in \eqref{dadef} comes from plugging these expressions into the conformal partial wave $(x_{12})^{-12}\mathcal{F}_T$. All of the integrals can be done analytically. We first convert the transverse directions to polar coordinates and do the transverse integrals. The result consists of rational functions of $\tau$, and rational functions multiplied by $\log (2 \pm  i \tau)$. The integral over $\tau$ is then done analytically. The result up to an overall positive numerical factor is \eqref{dgfinal}. The complete calculation of the integrals is provided in the Mathematica notebook included with the arXiv submission.

\section{Translation to transverse polarizations}\label{app:nicepol}

In section \ref{s:texchange}, we computed the correlator $\delta G$ with a particular choice of polarizations $\e$, $\overline{\e}$ given in \eqref{specialep}, and found a result of the form
\be\label{ggff}
-i \delta G = g_0 + g_2 \lambda^2 + g_4 \lambda^4 \ .
\ee
Symmetries only allow three different polarization tensor structures in the final answer, so these three coefficients contain the full information about the integrated correlator for arbitrary polarizations.  In this appendix we convert to a symmetric traceless tensor polarization $e_{ij}$ satisfying $e \cdot \hat{y} = 0$.  This is necessary in order to prove that $g_2 \geq 0$ (the other two constraints are obvious already in the form \eqref{ggff}, since we can set $\lambda = 0$ or $\lambda \to \infty$ to isolate those terms). 

The symmetries of the problem are identical to the energy correlator calculation of Hofman and Maldacena after a rotation by $\pi/2$ in the Euclidean $\tau y$-plane \cite{anec}, so the result must take the form
\be\label{dgts}
-i \delta G = C(e\cdot \ebar)\left[ 1 + \hat{t}_2 \left( \frac{e_{ij}\ebar^i_{\ k}n^j n^k}{e\cdot \ebar} - \frac{1}{3} \right) + \hat{t}_4\left( \frac{(e_{ij}n^i n^j)(\ebar_{kl}n^k n^l)}{e\cdot \ebar} - \frac{2}{15}\right) \right]\ ,
\ee
where $e\cdot \ebar = e_{ij} \cdot \ebar^{ij}$,  $n = i\hat{t}$. This also contains the full information about arbitrary polarizations, so we can translate between the two by converting this to general polarizations using the projector onto transverse traceless tensors, then contracting with $\e$, $\overline{\e}$.  For example, with $\mu$ running over all coordinates and $i$ running over $(t, \vec{x})$, the first term is converted as 
\begin{align}
e_{ij}\ebar^{ij} &\to e_{\alpha\beta}\ebar_{\gamma\delta}\left[ \half ( \delta^{\alpha \gamma} - n^\alpha n^\gamma)(\delta^{\beta\delta} - n^\beta n^\delta)  + (\alpha\leftrightarrow\beta)- \frac{1}{3}(\delta^{\alpha\beta} - n^\alpha n^\beta)(\delta^{\gamma\delta} - n^\gamma n^\delta) \right] \nn\\
&\to (\e . \overline{\e} - \e . n \overline{\e}. n)^2 - \frac{1}{3}(\e.n)^2(\overline{\e}.n)^2 \\
&= \frac{2}{3} + 4\lambda^2 + 4 \lambda^4  \ . \nn
\end{align}
Following the same procedure for the other structures gives the mapping between \eqref{dgts} and \eqref{ggff}:
\begin{align}
g_0 &= \frac{4}{9}C\left( \frac{3}{2} + \frac{ \hat{t}_2}{2} + \frac{4\hat{t}_4}{5}\right)\\
g_2 &= 2 C\left(2 + \frac{\hat{t}_2}{3} - \frac{4 \hat{t}_4}{15} \right) \nn\\
g_4 &= 4C\left(1 - \frac{ \hat{t}_2}{3} - \frac{2 \hat{t}_4}{15}\right)\nn
\end{align}
Comparing to \cite{Hofman:2008ar} (equation (2.38)), we see that the three coefficients $g_{0,2,4}$ correspond to the three combinations in \eqref{dgts} that must be positive in order to have $i \delta G \geq 0$ for arbitrary polarizations. This mapping is also useful to compare our expressions to the notation used in the calculations of bulk time delays \cite{Camanho:2014apa}.

\section{Causality violation from non-conserved spin-2 exchange}\label{app:spin8}

In this appendix we illustrate why stress tensor exchange cannot be cancelled by a non-conserved spin-2 operator in the specific case of  $\Delta_X = 8$. (An identical story applies to $\Delta_X = 6$.) In $d=4$, $\langle TTT\rangle $ has three allowed structures, but $\langle TTX\rangle$ has only two. The corresponding 3-point coupling constants will be denoted $\beta_1$ and $\beta_2$. In addition to modifying the allowed structures, there is also explicit $\Delta_X$ dependence in the conformal block $\bF_X$, which introduces constant coefficients as well as factors of $(\bz/z)^{\Delta_X}$.  Adding the $T$-block from \eqref{dgfinal} to the $X$-block, the total leading contribution in the Regge-bulk-point limit takes the form
\small
\begin{align}
\delta G \sim \frac{i}{(1-\rho)^5} \left(
\begin{matrix}
 t_4 + c_X \beta_1\\ -4 \left(  t_4 + c_X \beta_1 \right) \\   t_4 + c_X \beta_1 
\end{matrix}\right)
 + \frac{i}{(1-\rho)^4} \left(\begin{matrix}
 t_4 + c_X (\beta_1 + a_1\beta_2)\\
 -4 \left[t_4 + c_X(\beta_1 +a_2 \beta_2)\right]\\
 t_4 + c_X \beta_1
 \end{matrix}\right)
  + \cdots
\end{align}
\normalsize
We have organized the constrained combinations multiplying $\lambda^0,\lambda^2,$ and $\lambda^4$ into a column vector, and defined $c_X \sim c_{\psi\psi X}/c_{\psi\psi T}$ up to a positive constant. The $a_i$ are constants. Note that the leading term in $\bF_X$ is identical to the leading term in $\bF_T$, up to a constant. This is because $(z/\bz)^{\Delta_X} \to 1$ in the Regge-bulk-point limit. Now we apply the chaos bound. The leading term is causal only if it vanishes, so set $c_X = -t_4/\beta_1$ and proceed to the subleading term.  This sets $\beta_2 = 0$. Finally, the order $(1-\rho)^3$ terms (not written) can be expressed in terms of just the stress tensor couplings, $t_2$ and $t_4$, and requiring these terms to be positive sets $t_2 = t_4 = 0$.  

We should also consider sums of many different spin-2 operators $X_i$, each with its own set of couplings. In this case, with enough parameters it may be possible to cancel the causality violating terms in $\bF_T$, and it seems to rule this out would require an analysis of the stress-tensor 4-point function, $\langle TTTT\rangle$, as discussed in section \ref{s:restored}. 

\end{spacing}

\end{document}